\documentclass[twocolumn,aps,prd,amsmath,amssymb,groupedaddress,showpacs]{revtex4-1}

\usepackage{graphicx}

\newcommand{\beq}{\begin{eqnarray}}
 \newcommand{\eeq}{\end{eqnarray}}
\newcommand{\be}{\begin{equation}}
 \newcommand{\ee}{\end{equation}}

\def\fun#1#2{\lower3.6pt\vbox{\baselineskip0pt\lineskip.9pt
\ialign{$\mathsurround=0pt#1\hfil ##\hfil$\crcr#2\crcr\sim\crcr}}}

\newcommand{{\SD}}{\rm SD}
\newcommand{{\Lc}}{\mathcal{L}}
\newcommand{{\Mc}}{\mathcal{M}}

\newcommand{\vex}{\mbox{\boldmath${\rm x}$}}
\newcommand{\vey}{\mbox{\boldmath${\rm y}$}}

\newcommand{\ver}{\mbox{\boldmath${\rm r}$}}
\newcommand{\vesig}{\mbox{\boldmath${\rm \sigma}$}}

\newcommand{\veP}{\mbox{\boldmath${\rm P}$}}
\newcommand{\vep}{\mbox{\boldmath${\rm p}$}}

\newcommand{\vez}{\mbox{\boldmath${\rm z}$}}
\newcommand{\veA}{\mbox{\boldmath${\rm A}$}}

\newcommand{\veL}{\mbox{\boldmath${\rm L}$}}

\newcommand{\veR}{\mbox{\boldmath${\rm R}$}}

\newcommand{\vepi}{\mbox{\boldmath${\rm \pi}$}}
\newcommand{\veta}{\mbox{\boldmath${\rm \eta}$}}
\newcommand{\veB}{\mbox{\boldmath${\rm B}$}}
\newcommand{\veH}{\mbox{\boldmath${\rm H}$}}
\newcommand{\veE}{\mbox{\boldmath${\rm E}$}}

\newcommand{\veF}{\mbox{\boldmath${\rm F}$}}

\newcommand{\llan}{\langle\langle}
\newcommand{\rran}{\rangle\rangle}
\newcommand{\lan}{\langle}
\newcommand{\ran}{\rangle}

\begin{document}

\title{Meson Spectrum in Strong Magnetic Fields} 

\author{M. A. Andreichikov}
\email{andreichicov@mail.ru}
\affiliation{Institute of Theoretical and Experimental
Physics\\ 117118, Moscow, B.Cheremushkinskaya 25, Russia}
\affiliation{Moscow Institute of Physics and Technology,\\Moscow region, 141700
 Russia}
\author{B. O. Kerbikov}
\email{borisk@itep.ru}
\affiliation{Institute of Theoretical and Experimental
Physics\\ 117118, Moscow, B.Cheremushkinskaya 25, Russia}
\affiliation{Moscow Institute of Physics and Technology,\\Moscow region, 141700
 Russia}
\author{V. D. Orlovsky}
\email{orlovskii@itep.ru}
\affiliation{Institute of Theoretical and Experimental
Physics\\ 117118, Moscow, B.Cheremushkinskaya 25, Russia}
\author{Yu. A. Simonov}
\email{simonov@itep.ru}
\affiliation{Institute of Theoretical and Experimental
Physics\\ 117118, Moscow, B.Cheremushkinskaya 25, Russia}

\date{\today}

\begin{abstract}
We study the relativistic quark-antiquark system  embedded in  magnetic field
(MF). The  Hamiltonian containing confinement, one gluon exchange and spin-spin
interaction is derived. We analytically follow the evolution of the lowest
meson states as a functions of MF strength. Calculating the one gluon exchange interaction energy $\lan V_{\rm OGE}\ran$ and spin-spin contribution $\lan a_{SS} \ran$ we have observed, that these corrections remain finite at large MF, preventing the vanishing of the total $\rho$ meson mass at some $B_{\textrm{crit}}$, as previously thought. We display the $\rho$ masses as functions of MF in comparison with recent lattice data.
\end{abstract}


\maketitle

\section{Introduction}

During the last years we have witnessed an impressive progress of the
fundamental physics in ultra-intense magnetic field (MF) reaching the strength
up to $eB \sim 10^{18} G \sim m^2_\pi $ \cite{1}. Until  recently magnetars
\cite{2} were the only physical objects, where such, or somewhat weaker MF
could be realized. Now MF of the  above  strength and even stronger is within
reach in peripheral  heavy ion  collisions at RHIC and LHC \cite{3}. High
intensity lasers is another perspective tool to achieve MF beyond the Schwinger
limit \cite{4}. On the theoretical side  a striking progress has been achieved
along several lines. It is beyond our scope to discuss these works  or even
present a list of corresponding references. We mention only two lines of
research which have a certain overlap with our work. The first one \cite{5,6}
is the behavior of the hydrogen atom and positronium in very strong MF. The
second one \cite{7} is the conjecture of the vacuum reconstruction   due to
 vector meson  condensation in large MF. The relation between the  above studies and our
work will be clarified in what follows.

Our goal is  to study from the   first principles  the spectrum of a meson
composed of quark-antiquark embedded in MF. Use will be made of
Fock-Feynman-Schwinger representation  (see \cite{8} for review and references)
of the quark Green's function with strong (QCD) interaction and MF included. An
alternative approach could have been Bethe-Salpeter type formalism. However,
for  the confinement originating from the area law of the Wilson loop, the  use
of the gluon propagator is inadequate. Numerous attempts in this direction
failed because of gauge dependence and the vector character of the gluon
propagator, while confinement is scalar and gauge invariant. Therefore it is
sensible to use the path integral technique for QCD$+$QED Green's functions.
This method based on the proper-time formalism allows to represent the quark-antiquark Green's function via the Hamiltonian (see \cite{8*}  for a new derivation), and it was used in \cite{10} to construct explicit expressions for meson
Hamiltonians without MF. In this way spectra of light-light,
light-heavy and heavy-heavy mesons were computed with a good accuracy, using
the string tension $\sigma$, strong coupling constant $\alpha_S$ and quark
current masses as an input \cite{11},\cite{I}.

In what follows we expand this technique  to incorporate the effects of MF on
mesons. The latter contains: 1)  direct influence of MF on quark and antiquark,
and 2) the influence on gluonic fields, e.g., on the gluon propagator via $q\bar{q}$ loops
and on the gluon field correlators determining the string tension $\sigma$.
Since MF acts on charged objects, its influence on the gluonic degrees
of freedom enters only via $(N_c)^{-k}, ~~ k=1,2,...$, however, corrections of the second type can be important, as shown in \cite{12*}. 3) As was shown recently in the framework of our method, MF also changes quark condensate $\lan \bar q q\ran$ and quark decay
constants $f_\pi$ etc., and in this way strongly influences chiral dynamics \cite{12**}.

The important step in our relativistic formalism is the implementation of the
pseudomomentum notion and  center-of-mass (c.m.) factorization in MF, suggested in the
nonrelativistic case in \cite{12} for neutral two particle systems. Recently the c.m. factorization was proved for the neutral 3-body system in \cite{13}, the situation with charged 2-body system was clarified and an approximation scheme was suggested in \cite{13*}.

The plan of the paper is the following. Section 2 contains a brief pedagogical
reminder of how the two-body problem in MF is solved in quantum mechanics. The
central point  here is the integral of motion (``pseudomomentum'') which allows
the separation of the center of mass. Here we also  show how to diagonalize the
spin-dependent interaction. In section 3 we formulate the path integral for
quark-antiquark system with QCD$+$QED interaction. Then from Green's function
the relativistic Hamiltonian is obtained. Section 4 is devoted to the treatment
of confining and color Coulomb terms. Here we also present the derivation of the
 eigenvalue equations for the relativistic Coulomb problem.
 In section 5 we discuss the spectrum
of the system focusing on the regime of ultra-strong MF.  Section 6 contains
the discussion of the results,  comparison with lattice calculations, drawing
further perspectives and intersections of our results with those of other
authors \cite{5,6,7}. 

\section{Pseudomomentum and Wavefunction Factorization}

The total momentum of $N$ mutually interacting particles with translation
invariant interaction is a constant of motion and the center of mass motions
can be separated in Schroedinger equation. It was shown \cite{12} that a system
embedded in a constant MF also possesses a constant of motion
--``pseudomomentum''. As a result for the case  of zero total electric charge
$Q=0$ the c.m. motion  can be removed from the total Hamiltonian.
The simplest example is a two-particle system with equal masses $m_1=m_2=m$ and
electric charges $e_1 = - e_2=e$. We define \be \veR = \frac{\ver_1+\ver_2}{2},
~~ \veta= \ver_1 -\ver_2, ~~ \veP = \vep_1 +\vep_2.\label{1}\ee

Straightforward calculation  in the London gauge $\veA =\frac12 (\veB\times
\ver)$ yields
\begin{multline}
 \hat H =\frac{1}{4m} \left(\veP-\frac{e}{2} \left(\veB\times
\veta\right)\right)^2+ \\+ \frac{1}{m} \left(-i\frac{\partial}{\partial
\veta}-\frac{e}{2}(\veB\times \veR)\right)^2+V(\eta).\label{2}
\end{multline}
One can verify that the following ``pseudomomentum'' operator $\veF$ commutes
with the Hamiltonian (\ref{2}) \be \hat{\veF} =\veP +\frac{e}{2}(\veB\times
\veta).\label{3}\ee This immediately leads to the following factorization of
the wave function (WF) \be \Psi (\veR,\veta) =\varphi (\veta) \exp \left\{
i\veP \veR -i\frac{e}{2}(\veB\times \veta)\veR\right\}.\label{4}\ee

For the oscillator-type potential $V(\eta)$ the problem reduces to a set of
three oscillators, two of them are in a plane perpendicular to the magnetic
field and their frequencies are degenerate, while  the third  one is connected
solely with $V(\eta)$.

Next we briefly elucidate
 the spin interaction in presence of MF. The corresponding part of
 the Hamiltonian may be written as
 \be \hat H_s = a_{hf} (\vesig_1 \vesig_2) -\mu\veB (\vesig_1-\vesig_2),\label{5}\ee
where $e_1=-e_2=e>0$ and $\mu>0$. Diagonalization of $\hat H_s$ yields the
following four eigenvalues  e.g. for $u\bar u$ system, comprising both $\rho$
and $\pi$ levels. \be E_{1,2}^{(s)} =a_{hf}, ~~ E_{3,4}^{(s)} =\pm a_{hf}
\left( 2\sqrt{1+\left(\frac{\mu B}{a_{hf}}\right)^2}\mp 1\right),\label{6}\ee
where we assume that $\veB$ is aligned along the positive $z$-axis and
$B=|\veB|$. In a strong MF  when $\mu B
> a_{hf}$ spin-spin interaction becomes unimportant and $E_{3,4}^{(s)} \simeq \pm
2\mu B$. For the lowest level $E_4^{(s)}$ this corresponds to a configuration
$|+-\ran$ when the spin of negatively charged particle is aligned antiparallel
to $\veB$, and the spin of the positively charged one -- parallel to $\veB$.
This means that the spin (and isospin) are no more good quantum numbers and
eigenvalues (\ref{6}) correspond to the mixture of spin 1 and spin  0 states.
As a result the $q \bar q$ state will split into 4 states (two of them
coinciding $E_1^{(s)} = E_2^{(s)}$).
 Till now
we treated a nonrelativistic system, to incorporate relativistic effects we
shall exploit the path integral form of relativistic Green's functions
\cite{8,8*}.

\section{Relativistic $q\bar q$ Green's function and effective Hamiltonian}

The derivation of the relativistic Hamiltonian of the $q\bar{q}$ system in MF consist of several steps. The first one is the 4d relativistic path integral for the $q\bar{q}$ Green's function. The starting point is the Fock-Feynmann-Schwinger (world-line) representation
of the quark Green's function \cite{8}. The role of the ``time'' parameter
along  the path $z^{(i)}_\mu(s_i)$ of the $i$-th quark is played by the
Fock-Schwinger proper time $s_i, i=1,2.$ Consider a quark with a charge $e_i$
in a gluonic field $A_\mu$ and  the electromagnetic vector potential
$A_\mu^{(e)}$, corresponding to a constant magnetic field $\textbf{B}$. Then the
quark propagator in the Euclidean space-time is \be S_i (x,y) = (m_i +\hat
\partial - i g \hat A - i e_i \hat A^{(e)})^{-1}_{xy} \equiv(m_i+\hat
D^{(i)})^{-1}_{xy}.\label{7}\ee

The path-integral representation for $S_i$ \cite{8} is
\begin{multline}
S_i (x,y) = (m_i -
\hat D^{(i)})\int^\infty_0 ds_i (D^4z)_{xy}e^{-K_i}\Phi_\sigma^{(i)} (x,y)\equiv \\ \equiv
(m_i-\hat D^{(i)}) G_i (x,y), \label{8}
\end{multline}
\begin{widetext}
where
\begin{gather}
K_i = m_i^2 s_i +
\frac14\int^{s_i}_0 d\tau_i
\left(\frac{dz_\mu^{(i)}}{d\tau_i}\right)^2,\label{9} \\
\Phi^{(i)}_\sigma
(x,y) =P_AP_F \exp \left( ig \int^x_y A_\mu dz_\mu^{(i)}+ ie_i \int^x_y
A^{(e)}_\mu dz_\mu^{(i)}\right) \exp \left( \int^{s_i}_0
d\tau_i \sigma_{\mu\nu} (gF_{\mu\nu} + e_i B_{\mu\nu})\right).\label{10}
\end{gather}
Here $F_{\mu\nu} $ and $B_{\mu\nu}$ are correspondingly gluon and MF tensors,
$P_A, P_F$ are ordering operators,  $\sigma_{\mu\nu} = \frac{1}{4i}
(\gamma_\mu\gamma_\nu-\gamma_\nu\gamma_\mu)$. Eqs. (\ref{7}-\ref{10}) hold for
the quark, $i=1$, while  for the antiquark one should reverse the signs of
$e_i$ and $g$. In explicit form one writes \be \sigma_{\mu\nu} F_{\mu\nu} =
\left(
\begin{array}{ll} \vesig\veH&\vesig \veE\\\vesig\veE& \vesig
\veH\end{array}\right),~~ \sigma_{\mu\nu}B_{\mu\nu} = \left(
\begin{array}{ll} \vesig\veB&0\\0&\vesig\veB\end{array}\right).\label{11}\ee

Next we consider $q_1\bar q_2$ system born at the point $x$ with the current
$j_{\Gamma_1}(x) = \bar q_1 (x) \Gamma_1 q_2 (x)$ and  annihilated at the point
$y$ with the current $j_{\Gamma_2} (y)$. Here $x$ and $y$ denote the sets of
initial and final coordinates of quark and antiquark.
 Using the nonabelian Stokes theorem
and cluster expansion for the gluon field(see \cite{11} for reviews) and
leaving the MF term intact, we can write
\begin{multline}
G_{q_1\bar q_2} (x,y) = \int^\infty_0 ds_1 \int^\infty_0 ds_2
(D^4z^{(1)})_{xy} (D^4z^{(2)})_{xy} e^{-K_1-K_2}\textrm{tr} \lan  \hat TW_\sigma (A)\ran_A\times \\
\times \exp \left(ie_1 \int^x_y A^{(e)}_\mu dz^{(1)}_\mu -ie_2 \int^x_y A^{(e)}_\mu
dz^{(2)}_\mu +e_1\int^{s_1}_0 d\tau_1 (\vesig \veB) -e_2\int^{s_2}_0 d\tau_2
(\vesig \veB)\right),\label{12}
\end{multline}
where \be \hat T = \Gamma_1 (m_1  -\hat D_1)
\Gamma_2 (m_2 -\hat D_2),\label{13}\ee and $\Gamma_1 = \gamma_{\mu},\ \Gamma_2
= \gamma_{\nu}$ for vector currents, $\Gamma_i=\gamma_5$ for pseudoscalar currents, while \be \lan W_\sigma (A)\ran_A = \exp
\left( -\frac{g^2}{2} \int d \pi_{\mu\nu} (1) d\pi_{\lambda\sigma} (2) \langle
F_{\mu\nu}(1)F_{\lambda\sigma}(2)  \rangle + \mathcal{O}(\langle FFF \rangle) \right),\label{14}\ee
\end{widetext}
where $d\pi_{\mu\nu} \equiv ds_{\mu\nu} + \sigma_{\mu\nu}^{(1)} d{\tau_1} -
\sigma_{\mu\nu}^{(2)} d\tau_2,$ and $ds_{\mu\nu}$ is an area element of the
minimal surface, which can be constructed using straight lines, connecting the
points $z_i^{(1)} (t)$ and $z_j^{(2)} (t)$ on the paths of $q_1$ and $\bar
q_2$ at the  same time $t$ \cite{8, 10}. Note, that operator $\hat{T}$ actually do not participate in field averaging procedure: as was shown in \cite{13**}, the following replacement is valid: $m-\hat{D} \rightarrow m- i\hat{p}, \, p_{\mu} = \frac{1}{2}\left( \frac{dz_\mu}{d\tau}\right)_{\tau=s}$.

As a result of the first step the $q\bar{q}$ Green's function is represented as a 4d path integral (including Euclidean time paths) and in addition also integrals over proper times $s_1, s_2$. In the second step one introduces monotonic Euclidean time $t_E(\tau) = x_4 + \frac{\tau}{s}T$, where $T\equiv |x_4-y_4|$, so that $z_4(\tau) = t_E(\tau)+\Delta z_4(\tau)$, where $\Delta z_4(\tau)$ is fluctuation of time trajectory around $t_E(\tau)$. This new variable $t_E$ is an ordering parameter for trajectories $\textbf{z}^{(1)}(t_E), \, \textbf{z}^{(2)}(t_E)$, and proper times transform into physical parameters -- virtual $q$ and $\bar{q}$ energies $\omega_i \equiv \frac{T}{2s_i}$, so that $ds_i = -\frac{T}{2\omega_i^2}d\omega_i$.

Combining for simplicity all fields into one Wilson loop $W (A, A^{(e)})$, one can rewrite the Green's function in new variables as
\begin{multline} \label{Green's_function}
G_{q_1\bar q_2}(x,y) = \frac{T}{8\pi}
\int^\infty_0 \frac{d\omega_1}{\omega_1^{3/2}} \frac{d\omega_2}{\omega_2^{3/2}}
(D^3z^{(1)} D^3z^{(2)} )_{\vex\vey} \times \\ \times e^{-K_1(\omega_1) - K_2(\omega_2)} \llan \hat{T}
W_F\rran_{\Delta z_4 },
\end{multline}
(see \cite{8*} for details of derivation).
Here $K_1(\omega_1), \, K_2(\omega_2)$ are obtained from $K_i$ in (\ref{9}) by the same replacement $\frac{dz_i}{d\tau_i} = 2\omega_i\frac{dz_i}{dt_E}$,
\begin{multline}
 K_1 (\omega_1)+ K_2 (\omega_2)  = \left( \frac{m_1^2+\omega_1^2}{2\omega_1}
+ \frac{m_2^2+\omega_2^2}{2\omega_2}\right) T + \\+ \int^T_0 d t_E \left[
\frac{\omega_1}{2} \left( \frac{d\vez^{(1)}}{dt_E}\right)^2 +\frac{\omega_2}{2}
\left( \frac{d\vez^{(2)}}{dt_E}\right)^2 \right].\label{K1+K2}
\end{multline}

The final step is the use of the Wilson loop dynamics to express all dynamics in terms of instantaneous interaction. Indeed, the quadratic field correlator in (\ref{14}) is represented through two scalar functions $D(z)$ and $D_1(z)$ (see, e.g., \cite{11,SS} for details), first of them is responsible for confinement, while the second one gives one gluon exchange (OGE) potential.
%
%
So, for the case of zero quark orbital momenta with the minimal
surface, discussed above, integrating over relative time $\nu = t_E^1-t_E^2$ in $D(\nu, \textbf{z}_1-\textbf{z}_2), \, D_1(\nu, \textbf{z}_1-\textbf{z}_2)$ one obtains a simple instantaneous answer for spin-independent (SI) part of $\langle
W_{\sigma}(A)\rangle_{A}$,
\begin{widetext}
\be
\langle W_{\sigma}(A) \rangle_{A}^{SI} = \exp \left(
- \int\limits^{T}_{0} d t^E \left[ \sigma|\vez^{(1)} - \vez^{(2)}| -  \frac{4}{3}
\frac{\alpha_s}{|\vez^{(1)} - \vez^{(2)}|}  \right] \right),\label{15a}
\ee
containing $V_{\rm conf}(r) = \sigma r$ and $V_{\rm OGE}(r) = -\frac{4\alpha_s}{3r}$. Here $\sigma$ is
the  QCD string tension, $\sigma = 0.18 \ \mathrm{GeV}^2$ in our calculations.

First we need to find the Hamiltonian $H_{q_1 \bar q_2}$ of the system at
$t_1^E = t_2^E = t^E$. To this end we define the Euclidean Lagrangian $L^E_{q_1
\bar q_2} $. We write $\frac{d z^{(i)}}{d\tau_i} =
2\omega_i\frac{dz^{(i)}_k}{dt^E} = 2\omega_i \dot{z}_k,\ k = 1,2,3$. Then all
terms in the exponents in (\ref{12}), (\ref{14}) and (\ref{15a}) can be
represented as $\exp(-\int dt^E L_{q_1 \bar q_2}^E)$ and thus we arrive at the
following representation:
\be
G_{q_1\bar{q}_2}(x,y) = \frac{T}{8\pi} \int\limits_0^\infty \int\limits_0^\infty \frac{d\omega_1 d\omega_2}{(\omega_1\omega_2)^{3/2}} (D^3z^{(1)}D^3z^{(2)})_{\textbf{xy}} \, \textrm{tr}\left(e^{-S^E_{q_1\bar{q}_2}} \hat{T}\right)
\ee
with the action
\begin{multline}
 S^E_{q_1\bar q_2} = \int^{T}_0 dt^E \left[
 \sum_i \left( \frac{\omega_i}{2} (\dot z_k^{(i)})^2 - ie_i A_k^{(e)} \dot z^{(i)}_k  \right)+  \frac{\omega_1+\omega_2}{2} + \frac{m_1^2}{2\omega_1} +
 \frac{m_2^2}{2\omega_2} +  e_1 \frac{\vesig_1 \veB}{2\omega_1}+e_2 \frac{\vesig_2 \veB}{2\omega_2} + \right. \\ +   \left.\sigma |\vez^{(1)} -\vez^{(2)}| -\frac43
 \frac{\alpha_s}{|\vez^{(1)}-\vez^{(2)}|}\right].\label{16}
\end{multline}

 Here $A_k^{(e)}$ is the $k$--th component of the QED vector potential. The next step
 is the transition to the Minkowski metric. This is easy, since confinement is already expressed
 in terms of string tension. We have $\exp(-\int L^E dt_E) \rightarrow \exp(i \int L^M dt_M),\ t_E \rightarrow it_M$,
  and

\be \label{17}
  p^{(i)}_k =
 \frac{\partial L^M}{\partial\dot z_k^{(i)}} =\omega_i \dot z_k^{(i)} + e_i
 A_k^{(e)}, \quad H_{q_1\bar q_2} = \sum_i \dot z_k^{(i)} p^{(i)}_k -L_M
\ee
 Explicit expression for Hamiltonian without spin-dependent terms is
\be\label{Hamiltonian}
\quad H_{q_1\bar q_2} = \sum\limits_{i=1,2} \frac{(\textbf{p}^{(i)}-e_i\textbf{A}(\textbf{z}^{(i)}))^2 + m_i^2 +\omega_i^2 - e_i\vesig^{(i)}\textbf{B}}{2\omega_i} + \sigma |\textbf{z}^{(1)}-\textbf{z}^{(2)}| - \frac{4}{3}\frac{\alpha_s}{|\textbf{z}^{(1)}-\textbf{z}^{(2)}|}.
\ee

The $q \bar q$ Green's function (\ref{Green's_function}) takes the ``heat--kernel''
form, when going back to Euclidean time with Hamiltonian (\ref{Hamiltonian})
\be
G_{q_1 \bar q_2 }(x,y) = \frac{T}{8\pi}\int\limits_0^\infty \frac{d\omega_1}{\omega_1^{3/2}} \int\limits_0^\infty \frac{d\omega_2}{\omega_2^{3/2}} \left\langle \textbf{x}\left| \textrm{tr}(\hat{T} e^{-H_{q_1\bar{q}_2}T})\right|\textbf{y}\right\rangle.
\ee
The c.m. projection of the Green's function yields 
\be
\int G_{q_1 \bar q_2 }(x,y)d^3(x-y)= \frac{T}{8\pi}\int\limits_0^\infty \frac{d\omega_1}{\omega_1^{3/2}} \int\limits_0^\infty \frac{d\omega_2}{\omega_2^{3/2}} \sum\limits_{n=0}^{\infty} \varphi_n^2(0) \lan\textrm{tr}(\hat{T})\ran e^{-M_n(\omega_1, \omega_2)T},
\ee
\end{widetext}
where $\varphi_n$ and $M_n$ are eigenfunctions and eigenvalues of Hamiltonian $H_{q_1\bar{q}_2}$.
At large $T$ the integral over $\omega_1, \omega_2$ can be taken by the stationary point method, and hence the effective energies $\omega_i$ are to be found from the minimum of the total mass $M_n(\omega_1, \omega_2)$, as it was suggested in \cite{10}. To introduce the
minimization procedure and to check its accuracy we shall begin by the
calculation of the eigenvalues of one and two quarks in MF, and the energy of
the ground state of a relativistic charge in the atom  in the next section,
reproducing the known exact results.

We have the following equations defining $\omega_i$ from the total mass $M(\omega_i)$
\be
 \hat H \psi = M_n
 (\omega_1, \omega_2) \psi, ~~ \frac{\partial M_n(\omega_1,\omega_2)}{\partial \omega_i} =
 0.\label{19}
\ee

For a single quark in MF the first of the above equations gives \be M_n(\omega) =
\frac{p^2_z +m^2_q+ |eB| (2n+1) - eB\sigma_z}{2\omega} +
\frac{\omega}{2}.\label{20}\ee

Then the minimization over $\omega$ yields the correct answer \be \bar M_n = ( p^2_z+
m^2_q+ |eB| (2n+1) - eB \sigma_z)^{1/2}.\label{21}\ee
 Now we turn to the case
of $q_1 \bar q_2$ system and introduce the coordinates  which are the
generalization of (\ref{1}):

\be \veR = \frac{\omega_1 \vez^{(1)} + \omega_2 \vez^{(2)}}{\omega_1
+\omega_2},~~ \veta = \vez^{(1)} - \vez^{(2)},\label{22}\ee

\be \veP=-i\frac{\partial}{\partial\veR}, ~~ \vepi =-i
\frac{\partial}{\partial\veta}.\label{23}\ee

It is convenient to introduce the following two additional parameters
\beq
\tilde \omega = \frac{\omega_1 \omega_2}{\omega_1 + \omega_2},\ s =
\frac{\omega_1 - \omega_2}{\omega_1 + \omega_2}.
\eeq
Let us  consider the case of neutral meson, so that $e_1 = -e_2 = e$. Then the total
Hamiltonian may be written as \beq H_{q_1 \bar q_2} = H_B + H_{\sigma} + W,
\label{fh} \eeq where
\begin{multline}
H_B=\frac{1}{2\omega_1}\left[ \frac{\tilde \omega}{\omega_2} \veP +\vepi
-\frac{e}{2} \veB \times \left( \veR + \frac{\tilde \omega}{\omega_1}
\veta\right) \right]^2+ \\
  + \frac{1}{2\omega_2}\left[
\frac{\tilde \omega}{\omega_1} \veP -\vepi +\frac{e}{2} \veB \times \left( \veR
-\frac{\tilde \omega}{\omega_2} \veta\right) \right]^2 = \\ =
\frac{1}{2\tilde \omega}\left(\vepi -\frac{e}{2}\veB \times \veR + s
 \frac{e}{2}\veB \times \veta \right)^2 + \\+ \frac{1}{2(\omega_1 + \omega_2)}
 \left(\veP - \frac{e}{2}\veB \times \veta \right)^2.\label{a24}
\end{multline} 
 Equation (\ref{a24}) is an obvious generalization of (\ref{2}). The two other terms in (\ref{fh}) read
\be H_\sigma = \frac{m^2_1 + \omega^2_1 - e\vesig_1 \veB}{2\omega_1} +
\frac{m_2^2 +\omega_2^2 + e \vesig_2 \veB}{2\omega_2},\label{26}\ee \be
W=V_{\rm conf} + V_{\rm OGE} + \Delta W = \sigma \eta -\frac43
\frac{\alpha_s(\eta)}{\eta} +  \Delta W,\label{27}\ee and $\Delta W$
 contains self--energy and spin--spin contributions, which come from unaccounted spin-dependent terms of $\langle W_\sigma (A) \rangle$. One can verify, that the ``pseudomomentum'' operator in (\ref{3}), introduced in section II, commutes
  with $H_B$ and hence we can again separate the c.m. motion according to the
  ansatz (\ref{4}):
  \be
  H_B \Psi(\textbf{R},\veta) = \exp\left\{i\textbf{PR} - i\frac{e}{2}(\textbf{B}\times\veta)\textbf{R} \right\}\tilde{H}_B \varphi(\veta).
  \ee
   Then the problem reduces to the eigenvalue problem for
  $\varphi(\veta)$ with the Hamiltonian $\tilde{H}_B$ having the following form:
  \begin{multline}
  \tilde{H}_B = \frac{1}{2\tilde \omega}\left(-i \frac{\partial}{\partial \veta}
  + s \frac{e}{2}\veB \times \veta  \right)^2 + \\ + \frac{1}{2(\omega_1 + \omega_2)}
  \left(\veP - e\veB \times \veta \right)^2
  \end{multline}
For $\veP \times \veB = 0$ the system
   has a rotational symmetry and the c.m. is freely moving along the $z$-axis. Here
    we shall consider a state with zero orbital momentum $(\veL_{\eta})_z =
    \left[ \veta \times \frac{\partial}{i\partial \veta}  \right]_z = 0$.
    As a result $\tilde{H}_B$ is replaced by a purely internal space operator
    \be H_0 = \frac{1}{2\tilde \omega} \left( - \frac{\partial^2}{\partial
     \veta^2}+\frac{e^2}{4} (\veB\times \veta)^2
     \right),\label{28}\ee

To test our method we put $W = 0$ and arrive at the equation
     \be (H_0 + h_\sigma) \varphi = M(\omega_1,\omega_2) \varphi.\label{29}\ee
      Consequent minimization of $M(\omega_1,\omega_2)$ in $\omega_1, \omega_2$,
       as in (\ref{21}), yields the expected answer for the two independent quarks,
\begin{multline}
 M=\sqrt{m_1^2 + eB (2n_1+1) - e\vesig_1
\veB}+ \\ +\sqrt{m_2^2 + eB (2n_2+1) + e\vesig_2 \veB}.\label{30}
\end{multline}

We turn now to the particular case of charged two-body system in MF, $e_1=e_2=e$ and also $m_1=m_2$, when exact factorization of $\veR$ and $\veta$ can be done. In this case, for $\omega_1 = \omega_2 = \omega$ and $\textbf{P}\times \textbf{B} =0 $, the Hamiltonian has the following form \cite{8*}

\begin{multline}
 H_{q_1\bar{q}_2} = \frac{\textbf{P}^2}{4\omega}  + \frac{e^2}{4\omega} (\veB\times \veR)^2 + \frac{\vepi^2}{\omega} + \frac{e^2}{16
\omega} (\veB\times \veta)^2+  \\ + \frac{2m^2 + 2 \omega^2 - e (\vesig_1 +
\vesig_2) \veB}{2\omega}+ \frac{\sigma}{2} \left(\frac{\eta^2}{\gamma} + 
\gamma\right) +\\ + V_{\rm OGE} + V_{SS} + \Delta M_{SE}. \label{Hamiltonian_charged}
\end{multline}

\section{Treating confinement and gluon exchange terms. The absence of the magnetic QCD collapse}

From{(\ref{27}), (\ref{28}) it is clear, that inclusion of $V_{\rm conf}$ and
$V_{\rm OGE} $ in $H_0 + W $ leads to a differential equation in variables
$\eta_\bot, \eta_z,$ which can be solved numerically. However, in order to
obtain a clear physical picture, we shall represent $V_{\rm conf}$ in a
quadratic form. This will allow to get an exact analytic solution in terms of
oscillator functions with eigenvalue  accuracy of the order of $5\%$. The OGE contribution will be estimated as an average $\lan \varphi |V_{\rm
OGE}|\varphi\ran$, thus yielding an upper limit for the total mass.

For $V_{\rm conf}$ we choose the form \be V_{\rm conf} \to \tilde V_{\rm conf}
= \frac{\sigma}{2} \left( \frac{\eta^2}{\gamma} + \gamma\right)\label{31}\ee

Here $\gamma$ is a positive variational parameter; minimizing $\tilde V_{\rm
conf}$ w.r.t. $\gamma$, one returns to  $V_{\rm conf}$. We shall determine
$M(\omega_1 \omega_2, \gamma) $  corresponding to $\tilde V_{\rm conf}$, and to
define $\gamma$ an additional condition \be \left.\frac{\partial M (\omega_1,
\omega_2, \gamma)}{\partial \gamma}\right|_{\gamma=\gamma_0} =0\label{32}\ee
will be added to (\ref{19}). As a result  $M(\omega_1^{(0)}, \omega_2^{(0)},
\gamma_0)$ will be the final answer for the mass of the system, neglecting the $\Delta W$ contribution. The difference
of the exact numerical solution from that obtained with the genuine potential
$V_{\rm conf}$ does not exceed $5 \%$. The solution of the equation $(H_0 +
\tilde V_{\rm conf}) \varphi = M(\omega_1 , \omega_2 , \gamma_0) \varphi$ for
the ground state is
\be
 \psi(\veta) = \frac{1}{\sqrt{\pi^{3/2} r^2_\bot r_0}}
\exp \left( -\frac{\eta^2_\bot}{2r^2_\bot} -
 \frac{\eta^2_{z}}{2r^2_0}\right),
 \label{33}
 \ee
 where $r_\bot = \sqrt{\frac{2}{eB}} \left( 1+ \frac{4\sigma\tilde
 \omega}{\gamma e^2B^2}\right)^{-1/4},~ r_0 = \left( \frac{\gamma}{\sigma
 \tilde \omega}\right)^{1/4}$. As we shall see below, for the lowest mass eigenvalue with $eB\gg \sigma$,  one has
$r_\bot \approx \sqrt{\frac{2}{eB}},~ r_0  \approx \frac{1}{\sqrt{\sigma}}$ and
the $(q_1\bar q_2)$ system acquires the
 form of an elongated ellipsoid. Similar quasi--one--dimensional picture was observed before for the hydrogen--like atoms
in strong MF \cite{5,6}. In such geometrical configuration $V_{\rm OGE}$
manifests itself in a peculiar way, again similar to what happens in hydrogen,
or positronium atoms, and as was shown in \cite{12*} in QCD the outcome is also similar to the case of QED, with the screening of the diverging effects.

We turn now to the OGE term to treat it in our formalism. As a starting point we present another
check of our approach, namely we shall obtain the ground state energy of two
relativistic particles with opposite charges without MF  interacting via the
Coulomb potential. The corresponding Hamiltonian reads $H=H_0 + H_\sigma -
\frac{\alpha}{\eta},$ then $H \phi = M\phi$, and  for $eB =0$ we have \be M=-
\frac{\tilde \omega \alpha^2}{2} + \frac{m^2_1+ \omega^2_1}{2\omega_1}
+\frac{m^2_2+ \omega^2_2}{2\omega_2} .\label{39}\ee

Minimizing in $\omega_1$ in the limit $m_2\gg m_1$ (the hydrogen atom), one
obtains \be M=m_1 \sqrt{1-\alpha^2}+ m_2,\label{40}\ee which coincides with the
known  eigenvalue of the Dirac equation.

 In our  $(q_1\bar q_2)$ case one can calculate the expectation value of $V_{\rm OGE} = - \frac43
\frac{\alpha_s(\eta)}{\eta} $ with the asymptotic freedom and IR
 saturation behaviour in $\vep$--space (see \cite{16} for a derivation and a short review)
\be \alpha_s (q) =
 \frac{4\pi}{\beta_0 \ln \left( \frac{q^2+M^2_B}{\Lambda^2_{QCD}}\right)},
 \label{34}\ee where $\beta_0 = \frac{11}{3}N_c-\frac{2}{3}n_f$,
$M_B$ is proportional to $\sqrt{\sigma}$,  $M_B\approx 1$ GeV \cite{16}.
  With the wavefunction (\ref{33}) the average value of
$V_{\rm OGE} $ takes form
 \begin{multline}
 \Delta M_{\rm OGE} \equiv \int V_{\rm OGE} (q) \tilde \psi^2 (\textbf{q})
 \frac{d^3q}{(2\pi)^3} = \\= - \frac{4}{3\pi} \int^\infty_0 \alpha_s (q) dq
 e^{-\frac{q^2 r^2_\bot}{4}} I\left[
 \frac{q^2(r^2_0-r^2_\bot)}{4}\right],\label{35}
 \end{multline}
 where $\psi^2 (\textbf{q})$ is the Fourier transform of squared wave function $\psi^2( \veta)$ and $I(a^2)= \int^{+1}_{-1} dx e^{-a^2x^2}$. Estimating the integral in
(\ref{35}), for $eB \gg \sigma$, i.e. for $r_0 \gg r_\bot$ one obtains for
massless quarks \be \Delta M_{\rm OGE} \approx - \frac{16\sqrt{\pi}}{3r_0
\beta_0} \ln\ln \frac{r^2_0}{r^2_\bot} \approx - \sqrt{\sigma}\ln\ln
\frac{eB}{\sigma}.\label{36}\ee

With $eB$ increasing the upper bound for the $q\bar q$ mass is boundlessly
decreasing. The exact eigenvalue should lie  even lower.

This situation is similar to the hydrogen atom case, where $\Delta M_{\textrm{Coul}}$ diverges as $-\ln^2eB$, and in this case $e^+e^-$ loop contribution to the photon line stabilizes the result (the ``screening effect'' \cite{6}, \cite{7}). In our case the $q\bar{q}$ loop contribution to the OGE term can by written in a similar way, adding to the gluon loop also the Lowest Landau Level (LLL) of the $q\bar{q}$ in the MF,
\begin{widetext}
\be
\tilde{V}_{\textrm{OGE}}(Q) =-\frac{16\pi\alpha_s^{(0)}}
{3\left[Q^2\left(1+\frac{\alpha_s^{(0)}}{4\pi} \frac{11}{3} N_c \ln
\frac{Q^2+M^2_B}{\mu_0^2}\right)+ \frac{\alpha_s^{(0)}n_f |e_q
B|}{\pi}\exp\left(\frac{-q^2_\bot}{2|e_q B|}\right)T\left(\frac{q^2_3}{4\sigma}
\right)\right]},\label{V(Q)}
\ee
\end{widetext}
where $T(z)=-\frac{\ln\left( \sqrt{1+z}+\sqrt{z}\right)}{\sqrt{z(z+1)}} + 1$.
Calculating now the average value of (\ref{V(Q)}),
\be \label{dM_OGE}
\Delta M_{\textrm{OGE}} = \langle \tilde{V}_{\textrm{OGE}} \rangle,
 \ee
one obtains saturation of $\Delta M_{\textrm{OGE}}$ at large $eB$, as shown in Fig. 1, eliminating in this way the possible ``Color Coulomb catastrophe'', discussed in the first version of this paper \cite{Simonov:2012if}.
\begin{figure}
  \center{\includegraphics[height=7.0cm]{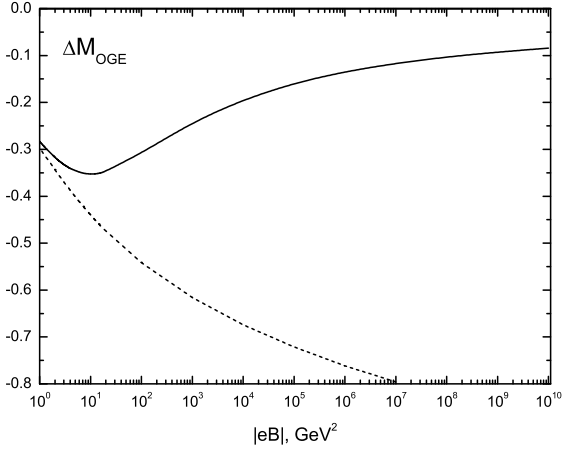}
    \caption{One gluon exchange correction to the meson mass  in GeV as a function of magnetic field with (solid line) (\ref{dM_OGE}) and without (broken line) (\ref{35}) account of quark loops contributions. \label{fig1}}}
\end{figure}

\section{Meson masses in magnetic field}

Our next task is to calculate analytically the mass $M_n (\omega_1, \omega_2,
\gamma)$  of a  $(q_1 \bar q_2)$ meson. We have to  solve the equation \be (H_0
+ H_\sigma +W) \Psi_n (\eta) = M_n(\omega_1, \omega_2, \gamma)\Psi_n (\eta),
\label{41}\ee where $H_0 ,H_\sigma,W$ are given in (\ref{26}), (\ref{27}), (\ref{28}), the total Hamiltonian for charged meson is given in (\ref{Hamiltonian_charged}).

The resulting mass for neutral meson without spin-dependent contribution from $\Delta W$ is
\begin{multline}
M_n(\omega_1, \omega_2, \gamma) = \varepsilon_{n_\bot , n_z} + \Delta M_{\rm OGE} +
\frac{m_1^2+\omega^2_1 - e\veB \vesig_1}{2\omega_1} + \\ +\frac{m_2^2+\omega^2_2 +
e\veB \vesig_2}{2\omega_2}  \equiv \bar{M}_n(\omega_1, \omega_2, \gamma) - \frac{e\veB \vesig_1}{2\omega_1} + \frac{e\veB \vesig_2}{2\omega_2},\label{42}
\end{multline}
where
\begin{multline}
\varepsilon_{n_\bot, n_z} =
\frac{1}{2\tilde \omega} \left[ \sqrt{ e^2 B^2 + \frac{4\sigma\tilde
\omega}{\gamma}} (2n_\bot +1)\right. +\\+ \left.\sqrt{\frac{4\sigma \tilde
\omega}{\gamma}}\left(n_z + \frac12\right)\right] + \frac{\gamma \sigma}{2},
\label{43}
\end{multline}
$\Delta M_{\rm OGE}$  is given by  (\ref{V(Q)}) and (\ref{dM_OGE}). So, for fixed $n$ we have four states for different quark spin orientations, $|++\rangle, |+-\rangle, |-+\rangle$ and $|--\rangle$, where $+/-$ are up/down directions of individual quark spins, with corresponding masses
\begin{gather}
M_n^{++}= \bar{M}_n - eB\left(\frac{1}{2\omega_1}-\frac{1}{2\omega_2}\right), \\ M_n^{--}= \bar{M}_n + eB\left(\frac{1}{2\omega_1}-\frac{1}{2\omega_2}\right), \\
M_n^{+-}= \bar{M}_n - eB\left(\frac{1}{2\omega_1}+\frac{1}{2\omega_2}\right), \\ M_n^{-+}= \bar{M}_n + eB\left(\frac{1}{2\omega_1}+\frac{1}{2\omega_2}\right).
\end{gather}

The spin-dependent part $\Delta W$ contains self-energy $V_{SE}$ and spin-spin $V_{SS}$ contributions. As was shown in \cite{spin_interactions}, the mass correction, corresponding to $V_{SE}$, is given by
\begin{multline}
\Delta M_{SE} = - \frac{3\sigma}{4\pi\omega_1}\left(1+\eta\left(\lambda\sqrt{2eB+m_1^2}\right)\right)- \\ - \frac{3\sigma}{4\pi\omega_2}\left(1+\eta\left(\lambda\sqrt{2eB+m_2^2}\right)\right),
\end{multline}
where $\eta(t) = t\int_0^\infty z^2K_1(tz)e^{-z}dz$ and $\lambda\sim 1$ GeV$^{-1}$ is vacuum correlation lengths.

Let us introduce now the spin-spin interaction. It has nondiagonal structure
\be \label{VSS}
V_{SS} = \frac{8\pi\alpha_s}{9\omega_1\omega_2}\delta^{(3)}(\textbf{r})\vesig_1\vesig_2 \equiv a_{SS} \vesig_1\vesig_2,
\ee
so we should diagonalize the Hamiltonian with respect to spin variables. This results in new four states, two of them are mixture of $|+-\rangle$ and $|-+\rangle$ states, corresponding to $\pi^0$ and $\rho^0$  with zero spin projection $s_z=0$, the other two states $|++\rangle$ and $|--\rangle$ correspond to $\rho^0$ states with $s_z=1$ and $s_z=-1$ (we consider ground state $n=0$). We note that both $V_{SS}$ and $\Delta M_{SE}$ are to be considered as
corrections and contain $\omega_1^{(0)}, \omega_2^{(0)}$, obtained from minimization of the remaining part of the meson mass. Note also, that masses $M_n^{+-}$ and $M_n^{-+}$ are symmetric with respect to $\omega_1\leftrightarrow \omega_2$ for equal quark masses, so in this case we have in fact only two variables in minimization procedure $\omega$ and $\gamma$.

The masses of first two states are:
\be\label{E1,2}
E_{1,2} = \frac{1}{2}(M_{11}+M_{22})\pm \sqrt{\left( \frac{M_{22}-M_{11}}{2}\right)^2+ 4a_{12}a_{21}},
\ee
where
\begin{gather}
M_{11} = \left.(M_0^{+-} + \Delta M_{SE}-\lan a_{SS} \ran)\right|_{\omega_1^{(0)}=\omega_2^{(0)}=\omega_\pm}, \nonumber \\ M_{22} = \left. (M_0^{-+} + \Delta M_{SE}-\lan a_{SS} \ran) \right|_{\omega_1^{(0)}=\omega_2^{(0)}=\omega_\mp},
\end{gather}
$a_{12}=a_{21} = \left.\lan a_{SS} \ran \right|_{\omega_1^{(0)}=\omega_\pm, \, \omega_2^{(0)}=\omega_\mp}$ and $\lan a_{SS} \ran$ is the averaging with the wave function (\ref{33}) (see \cite{spin_interactions} for derivation). The parameters $\omega_\pm$ and $\omega_\mp$ are obtained by minimizing of corresponding diagonal eigenvalues $M_0^{+-}$ and $M_0^{-+}$,  the parameter $\gamma_0$ (see (35)) for the ground state is defined from the condition
 \beq \label{gamma_condition}
 \left.\frac{\partial M_0(\omega_1,\omega_2,\gamma)
}{\partial \gamma}\right|_{\gamma=\gamma_0}=
 \left.\frac{\partial\varepsilon_{0,0} }{\partial
\gamma}\right|_{\gamma=\gamma_0} =0.
\eeq
It is easy to see, that at large $eB$ masses $E_{1,2}$ tend to diagonal values
\be
E_1(eB\rightarrow\infty) \rightarrow M_{22}, \quad E_2(eB\rightarrow\infty) \rightarrow M_{11}.
\ee

The remaining two states have masses
\be\label{E3,4}
E_3 = M_0^{++} + \Delta M_{SE}+ \lan a_{SS} \ran, \, E_4 = M_0^{--} + \Delta M_{SE}+ \lan a_{SS} \ran,
\ee
taken in point $(\omega_1^{(0)},\omega_2^{(0)},\gamma^0)$ in accordance with minimization conditions (\ref{19}) and (\ref{gamma_condition}).

 It should be noted, that actually we have eight states instead of four, since $q\bar{q}$ systems with different quark charges behave differently in MF, as we see from our Hamiltonians. Isospin is not conserved now and each neutral state splits into two states with different quark content $u\bar{u}$ and $d\bar{d}$.

Let us consider also the particular case of charged meson with Hamiltonian (\ref{Hamiltonian_charged}) in state with $s_z=1$ ($|++\ran$-state, corresponding to $\rho^+$). The eigenvalue, corresponding to this state, is given by the following expression
\begin{multline}\label{M_charged}
M_n(\omega,\gamma) = \frac{eB}{2\omega}(2N_\perp +1) + \sqrt{\left(\frac{eB}{2\omega}\right)^2  +\frac{2\sigma}{\omega\gamma}} (2n_\perp+1) + \\+ \sqrt{\frac{2\sigma}{\omega\gamma}} \left(n_\parallel+\frac{1}{2}\right)  - \frac{eB}{\omega} + \frac{\sigma\gamma}{2} + \frac{m^2+\omega^2}{\omega} +\\+ \Delta M_{OGE} +\Delta M_{SE} + \lan a_{SS} \ran.
\end{multline}

Among considered states, the mass of charged meson ground state ($\rho^+$ with $s_z=1$) and $E_2$, corresponding to $\pi^0$, tend to finite value at large MF due to cancellation of linearly growing terms in $\varepsilon_{n_\perp,n_z}$ and in $H_\sigma$, while other masses grow with $eB$. This is true, provided that the spin-spin contribution $\lan a_{SS} \ran$ remains finite at large MF. However, it contains the factor $\psi^2(0) \sim eB$, which leads to unbounded decrease of $E_2$. As was shown in \cite{spin_interactions}, this situation is not physical, the total mass eigenvalues should be positive, and the reason of this decrease is the unlawful use of the perturbation theory for the potential $c\delta^{(3)}(\textbf{r})$. One should replace $a_{SS}$ by a smeared out version, e.g.,
\be
\delta^{(3)}(\textbf{r}) \rightarrow \tilde{\delta}^{(3)}(\textbf{r}) = \left( \frac{1}{\lambda\sqrt{\pi}}\right)^3 e^{-\textbf{r}^2/\lambda^2}, \quad \lambda\sim 1~\textrm{GeV}^{-1}.
\ee
Using the wave function (\ref{33}), one obtains for $\lan a_{SS} \ran$
\be
\lan a_{SS} \ran = \frac{c}{\pi^{3/2}\sqrt{\lambda^2+r_0^2}(\lambda^2+r_\perp^2)}, \quad c=\frac{8\pi\alpha_s}{9\omega_1\omega_2}.
\ee
The smearing length $\lambda$ on the lattice corresponds to the lattice unit $a$ ($\lambda \sim a$), in physical situation the relativistic smearing is connected with the gluelamp mass parameters in $D(z)$ and $D_1(z)$, see \cite{SS} for details.

\begin{figure}[h]
  \centering
  \includegraphics[width=9cm]{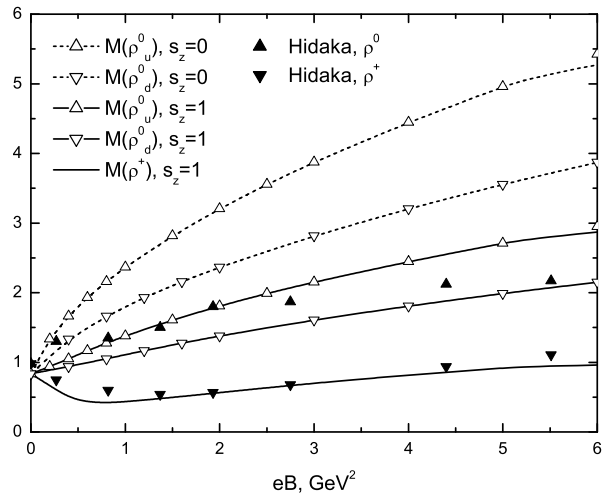}
  \caption{The masses of the systems in GeV as a functions of $eB$. See the text for explanations.}
\end{figure}

In Fig.2 we plot the masses of some selected systems as a functions of $eB$ ($e$ is the $\rho^+$ charge, not the charge of individual quarks).
Calculations were performed according to (\ref{E1,2}), (\ref{E3,4}), (\ref{M_charged}) and the minimization procedure.
The dashed curves correspond to the $\rho^0$ state with $s_z=0$ (eigenvalue $E_1$), the solid-symbol lines describe $\rho^0$ state with $s_z=1$, the lower solid curve refers to the state of charged meson $\rho^+$ with $s_z=1$. The black triangles are from lattice calculations \cite{18}. One
can see that the masses of first states are increasing, while the last one tends to finite limit in accordance with discussion above (note, that the results plotted in Fig.2 were obtained for massless quarks).

\section{Discussion and conclusions}

In our treatment of relativistic quark--aniquark system embedded in MF we
relied on pseudomomentum factorization of the wave function and relativistic
Hamiltonian technique. The Hamiltonian for mesons in MF, containing
confinement, one gluon exchange and spin interaction was derived. Using a suitable
approximation for confining force we were able to calculate analytically meson
masses as functions of the MF. In this paper to simplify things we started with $\rho^0$ meson states
 at $B = 0$ taking $\gamma_i$ in place of $\Gamma_1$ and $\Gamma_2$ in (\ref{13}). In this way we essentially
left aside the complicated problem of chiral dynamics and pseudo--Goldstone
spectrum. In this oversimplified picture the
lowest neutral state with $s_z=0$ is a mixture of the
$\rho^0$ and $\pi^0$, as can be seen from its spin and isospin structure.
Indeed, $u \bar u$ or $d\bar{d}$ system under consideration is a mixture of isospin $I = 0$
and $I=1$ states, and at large MF it has a spin structure $ |u \uparrow, \bar u
\downarrow \rangle$, which is a mixture of $S=0$ and $S=1$ states. We have calculated mass of the higher state of this mixture, which we call $\rho^0(eB)$, while the lower state, associated with $\pi^0(eB)$, can be subject to chiral corrections. These states have negative corrections from one-gluon exchange and spin-spin interactions. As was shown, these corrections (and the total mass) stay finite at large $B$, preventing the so called ``magnetic collapse in QCD'', discussed earlier in \cite{Simonov:2012if}.

 As shown in Fig. 2, our analytical results are in agreement with lattice calculations \cite{18} both
for $\rho^0$  and $\rho^+$ states. 

Another system which can be treated using the same
technique is the neutral 3--body system, like neutron. The results might be
important for the neutron stars physics.

The authors are grateful for useful discussions with V.A.~Novikov, A.E.~Shabad,
M.I.~Vysotsky and S.I.~Godunov. We  are  indebted  to  V.S.~Popov  for important
remarks.  B.K.  gratefully  acknowledge  the  support  RFBR  grant
10-02093111-NTSNIL-a. We are pleased to thank M.~Chernodub for his remarks and
suggestions in response to v.1 of \cite{Simonov:2012if}.

\end{document}